%% file: main.tex
\documentclass[sigconf]{acmart}
\acmSubmissionID{558}

\usepackage{booktabs} % For formal tables

\usepackage{tabularx}

\usepackage[ruled]{algorithm2e} % For algorithms

\SetAlFnt{\small}
\SetAlCapFnt{\small}
\SetAlCapNameFnt{\small}
\SetAlCapHSkip{0pt}

\copyrightyear{2024}
\acmYear{2024}
\setcopyright{rightsretained}
\acmConference[SIGGRAPH Conference Papers '24]{Special Interest Group on Computer Graphics and Interactive Techniques Conference Conference Papers '24}{July 27-August 1, 2024}{Denver, CO, USA}
\acmBooktitle{Special Interest Group on Computer Graphics and Interactive Techniques Conference Conference Papers '24 (SIGGRAPH Conference Papers '24), July 27-August 1, 2024, Denver, CO, USA}
\acmDOI{10.1145/3641519.3657437}
\acmISBN{979-8-4007-0525-0/24/07}

\begin{document}
% Title portion
\title{Strategy and Skill Learning for \\
Physics-based Table Tennis Animation}

\author{Jiashun Wang}
\email{jiashunw@cmu.edu}
\affiliation{%
 \institution{Carnegie Mellon University}
 \country{USA}}

\author{Jessica Hodgins}
\email{jkh@cmu.edu}
\affiliation{%
 \institution{Carnegie Mellon University \\
 and The AI Institute}
 \country{USA}}

\author{Jungdam Won}
\email{jungdam@imo.snu.ac.kr}
\affiliation{%
 \institution{Seoul National University}
 \country{South Korea}}
 
% DO NOT ENTER AUTHOR INFORMATION FOR ANONYMOUS TECHNICAL PAPER SUBMISSIONS TO SIGGRAPH 2019!
%\author{Gang Zhou}
%\orcid{1234-5678-9012-3456}
%\affiliation{%
%  \institution{College of William and Mary}
%  \streetaddress{104 Jamestown Rd}
%  \city{Williamsburg}
%  \state{VA}
%  \postcode{23185}
%\email{gang_zhou@wm.edu}
%\affiliation{%
%  \institution{Inria Paris-Rocquencourt}
%\email{beranger@inria.fr}
%\affiliation{%
% \institution{Rajiv Gandhi University}
%\email{aprna_patel@rguhs.ac.in}
%\affiliation{%
%  \institution{Tsinghua University}
%  \streetaddress{30 Shuangqing Rd}
%  \country{China}
%}
%\email{chan0345@tsinghua.edu.cn}
%\author{Ting Yan}
%\affiliation{%
%  \institution{Eaton Innovation Center}
%  \city{Prague}
%  \country{Czech Republic}}
%\email{yanting02@gmail.com}
%\author{Tian He}
%\affiliation{%
%  \institution{University of Virginia}
%  \department{School of Engineering}
%  \city{Charlottesville}
%  \state{VA}
%  \postcode{22903}
%  \country{USA}
%}
%\affiliation{%
%  \institution{University of Minnesota}
%\email{tinghe@uva.edu}
%\author{Chengdu Huang}
%\author{John A. Stankovic}
%\author{Tarek F. Abdelzaher}
%\affiliation{%
%  \institution{University of Virginia}
%  \department{School of Engineering}
%  \city{Charlottesville}
%  \state{VA}
%  \postcode{22903}
%  \country{USA}
%}

\renewcommand\shorttitle{Strategy and Skill Learning for Physics-based Table Tennis Animation}

\begin{abstract}
Recent advancements in physics-based character animation leverage deep learning to generate agile and natural motion, enabling characters to execute movements such as backflips, boxing, and tennis. However, reproducing the selection and use of diverse motor skills in dynamic environments to solve complex tasks, as humans do, still remains a challenge. We present a strategy and skill learning approach for physics-based table tennis animation. Our method addresses the issue of mode collapse, where the characters do not fully utilize the motor skills they need to perform to execute complex tasks. More specifically, we demonstrate a hierarchical control system for diversified skill learning and a strategy learning framework for effective decision-making. We showcase the efficacy of our method through comparative analysis with state-of-the-art methods, demonstrating its capabilities in executing various skills for table tennis. Our strategy learning framework is validated through both agent-agent interaction and human-agent interaction in Virtual Reality, handling both competitive and cooperative tasks.
\end{abstract}

\begin{CCSXML}
<ccs2012>
 <concept>
  <concept_id>10010520.10010553.10010562</concept_id>
  <concept_desc>Computer systems organization~Embedded systems</concept_desc>
  <concept_significance>500</concept_significance>
 </concept>
 <concept>
  <concept_id>10010520.10010575.10010755</concept_id>
  <concept_desc>Computer systems organization~Redundancy</concept_desc>
  <concept_significance>300</concept_significance>
 </concept>
 <concept>
  <concept_id>10010520.10010553.10010554</concept_id>
  <concept_desc>Computer systems organization~Robotics</concept_desc>
  <concept_significance>100</concept_significance>
 </concept>
 <concept>
  <concept_id>10003033.10003083.10003095</concept_id>
  <concept_desc>Networks~Network reliability</concept_desc>
  <concept_significance>100</concept_significance>
 </concept>
</ccs2012>
\end{CCSXML}

\ccsdesc[500]{Computing methodologies~ Physical simulation}
\ccsdesc{Computing methodologies~ Motion Processing}


\keywords{Character Animation, Physics-based Characters, Deep Reinforcement Learning, Multi-character Interaction, Virtual Reality, Table Tennis}

\begin{teaserfigure}
\includegraphics[width=\textwidth]{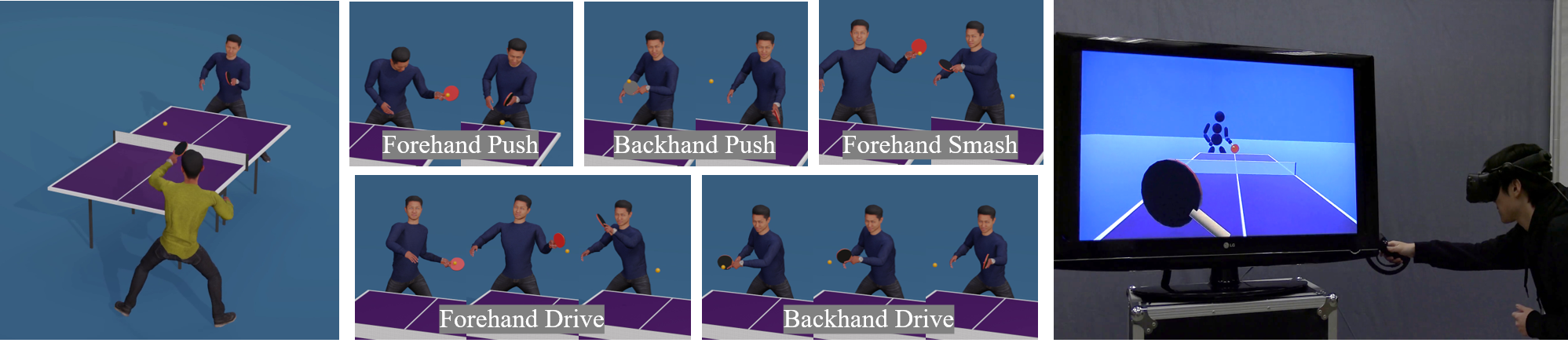}
\caption{On the left, two physically-simulated agents engage in a competitive match, controlled by our strategy and skill controllers. The center panels show the five learned skills: Forehand Drive, Push and Smash, Backhand Drive and Push, showcasing the skill controller's ability to execute a diverse set of skills. On the right, a human interacts with the simulated agent through VR.}
\label{fig:teaser}
\end{teaserfigure}

\maketitle

\input{Latex/1Introduction}

\input{Latex/2RelatedWork}
\input{Latex/3Overview}

\input{Latex/4SkillLearning}

\input{Latex/5StrategyLearning}
\input{Latex/6InteractionEnv}
\input{Latex/7Experiments}
\input{Latex/8Discussion}

\bibliographystyle{ACM-Reference-Format}
\bibliography{sample-bibliography}

\newpage
\input{Latex/FigureOnly}

\end{document}

% Title portion
\title{Appendix: Strategy and Skill Learning for \\
Physics-based Table Tennis Animation}

\author{Jiashun Wang}
\email{jiashunw@cs.cmu.edu}
\affiliation{%
 \institution{Carnegie Mellon University}
 \country{USA}}

\author{Jessica Hodgins}
\email{jkh@cmu.edu}
\affiliation{%
 \institution{Carnegie Mellon University \\
 and The AI Institute}
 \country{USA}}

\author{Jungdam Won}
\email{jungdam@imo.snu.ac.kr}
\affiliation{%
 \institution{Seoul National University}
 \country{South Korea}}

\renewcommand\shorttitle{Appendix: Strategy and Skill Learning for Physics-based Table Tennis Animation}
% \keywords{Character Animations, Motion Retargeting, }
\maketitle

\appendix

We describe the details of the states and actions used in our controllers, the training, and the implementation of the network architecture and hyper-parameters. We also provide more details of the skill and strategy evaluation.
\section{States and actions}
We describe the states and actions for both skill-level and strategy-level controllers.

\textit{Skill-level controller}. We follow~\cite{DBLP:journals/tog/PengGHLF22} to use the agent's local coordinate frame for the skill-level controller. The agent’s local coordinate frame is defined with the origin located at the root, the x-axis oriented along the root link’s facing direction, and the y-axis aligned with the global up vector. Agent's state $s\in \mathbb{R}^{226}$ consists of the height of the root, the rotation and position of the root in the local coordinate frame, the rotation and position of each joint in the local coordinate frame, the linear and angular velocity of each joint in the local coordinate frame, and the position of the paddle in the local coordinate frame. 
% \jungdam{what is the local coordinate actually? You mean the facing or root frame? Then, use the name of the frame instead of saying it as the local coordinate, which could be ambiguous.}
Ball state's $b \in \mathbb{R}^{9}$ includes the ball velocity, the distance between the ball and root, and the distance between the ball and paddle, computed in the agent’s local coordinate frame.
Target $y \in \mathbb{R}^{3}$ includes the distance between the ball and target in the agent's local coordinate frame. $\delta$ is a one-hot vector determining the skill to use.  Agent's action $a\in \mathbb{R}^{31}$ is the target joint angles except for the root for PD controllers and the blending weights $\varphi \in \mathbb{R}^{31}$ is used to mix the skill actions in a joint-wise manner.

\textit{Strategy-level controller}. Agent's state $s \in \mathbb{R}^{6}$ consists of the root position and paddle position in the table frame. Opponent's state $\tilde{s} \in \mathbb{R}^{6}$ consists of the root position and paddle position of the opponent in the table frame. Ball's state $b \in \mathbb{R}^{6}$ includes the ball position and velocity in the table frame. The table's longer edge is the x-axis, its shorter edge is the y-axis, and the z-axis is the gravity axis. Agent's strategy action $c \in \mathbb{R}^{6}$ includes the $\delta \in \mathbb{R}^{5}$ determining the skill to use and $y \in \mathbb{R}^{2}$ the target landing location on the table in the table frame. 
% \jungdam{Does the world coordinate frame match to a frame fixed to the table? If then, why don't you simply say 'a frame fixed to the table'?}

\section{Training details}
We utilize IsaacGym~\cite{DBLP:conf/nips/MakoviychukWGLS21} to train the agents with a simulation frequency of 120 Hz. For control policies, imitation policies run at 30Hz. Mixer and ball control policies run at 15Hz. The RL policies $\omega$ and $\pi$ are trained with proximal policy optimization (PPO)~\cite{DBLP:journals/corr/SchulmanWDRK17}. We collect 18 minutes of reference motion of a high-ranking player with the Vicon Motion Capture system. We perform a broad categorization including drive, push, and smash. 

All the networks are trained with Pytorch~\cite{DBLP:conf/nips/PaszkeGMLBCKLGA19} and all training processes are performed on NVIDIA RTX 4090. For the VR experiments, we finetune the policies to run with a control frequency of 60 Hz to enhance the interaction experience.  Each imitation policy is trained with 2 billion samples, taking 12 hours. Each ball control policy and mixer policy is trained with 4 billion samples, taking about 1 day. The strategy learning takes about 3 minutes to collect data and 5 minutes to train for one iteration. We apply 5 iterations for agent-agent experiments and 2 iterations for human-agent experiments. We split the reference motion into five subsets: Forehand Drive, Forehand Push, Forehand Smash, Backhand Drive and Backhand Push. A drive is a stroke that is primarily used to keep the ball in play with a fast and flat shot. A push is a stroke that requires the player to strike downwards on the back and underneath the ball to create a backspin.  A smash is a fast, hard and powerful stroke that drives the ball downward. We train imitation and ball control policies with each skill separately. We train the universal imitation policy and mixer policy with all the data.

\section{Network architecture and Hyper-parameters}

The imitation policies are modeled by a neural network to map to a Gaussian Distribution $\pi^{i}(a^i|s,z^i) = \mathcal{N} (\mu_{\pi^{i}}(s,z^i),\Sigma_{\pi^i})$, where $i\in\{1, 2, 3, 4, u\}$. Specifically, the mean $\mu_{\pi^{i}}$ is predicted by an MLP with three hidden layers of [1024, 1024, 512] units followed by a linear output layer, and the diagonal covariance matrix $\Sigma_{\pi^i}$ is set to $0.0025$ on each diagonal element. 
The value network is a similar architecture but the final output is a scalar. The encoder $q^{i}(z^i|s,s')$ and discriminator $D^{i}(s, s')$ are modeled by a single network which outputs both the mean of the encoder $\mu_{q^i}(s,s')$ and the discriminator value.  
The ball control policies $\omega^{i}(z^i|s,b,y) = \mathcal{N} (\mu_{\omega^{i}}(s,b,y),\Sigma_{\omega^i})$ are also modeled by a neural network to map to a Gaussian Distribution. Specifically, the mean $\mu_{\omega^{i}}(s,b,y)$ is predicted by an MLP with two hidden layers of [1024,512] units followed by a linear output layer, and the diagonal covariance matrix $\Sigma_{\omega^i}$ is set to $0.01$ on each diagonal element. The output $z^i$ is normalized with its norm before sending to the imitation policies. The mixer policy is similar to the ball control policy except it takes $s$, $b$, $y$ and $\delta$ as inputs and outputs $\mu_{\omega^{m}}(s,b,\delta,y)$ and blending weight $\varphi$. $f$ takes the strategy observation $o=(s, \tilde{s}, b)$ as input and outputs the strategy action $c=(\delta, y)$. It is modeled by an MLP with three hidden layers of [1024, 1024, 512] followed by a linear output layer. We use ReLU for all the activations except the outputs of the discriminator value and blend weight $\varphi$, which use a Sigmoid. Table~\ref{tab:hyper} reports the hyper-parameters used in our experiments.

\begin{table}
\caption{Hyper-parameters. }
\label{tab:hyper}
\vspace{-0.1in}
\begin{minipage}{\columnwidth}
\begin{center}
\begin{tabular}{|l|c|}
\hline
\textbf{Parameter} & \textbf{Value} \\ \hline
% Simulation rate (Hz)   & 240     \\ \hline
% Control rate (Hz)      & 30     \\ \hline
Discount factor $\gamma$        &  0.99     \\ \hline
GAE and TD $\lambda$  & 0.95     \\ \hline
Episode length & 500 \\ \hline
Learning rate         & $1.0e^{-5}$     \\ \hline
\# tuples per update      & 258944     \\ \hline
Policy Batch Size         & 16384     \\ \hline
Discriminator Batch Size $\gamma$        &  4096     \\ \hline
PPO Clip Threshold & 0.2 \\ \hline
Latent Space Dimension & 64     \\ \hline
Gradient Penalty Weight $\lambda_{gp}$      & 5     \\ \hline
Diversity Objective Weight $\lambda_{D}$      & 0.01         \\ \hline
CVAE KL Divergence Weight $\beta_{KL}$      & 0.01 \\ \hline
Skill Discovery Objective Weight $\beta$        &  0.5     \\ \hline
Paddle Reward Weight $w_{p}$      & 0.8    \\ \hline
Ball Reward Weight $w_{b}$  & 0.8         \\ \hline
Style Reward Weight $w_{r}$     &  0.2     \\ \hline

\end{tabular}
\end{center}
\end{minipage}
\end{table}%

\section{Skill evaluation}
ASE~\cite{DBLP:journals/tog/PengGHLF22}, CASE~\cite{DBLP:conf/siggrapha/DouCFKW23}, and ET are trained under the same setting as our mixer policy. ET learns a high-level policy to output the latent action for the universal imitation policy. For ET, we train the five skills first and then we fix the individual skill controllers and let the ET controller take control during the transition (when the ball passes the net until it is returned to the agent). For other time steps, we directly let each individual controller control the agent. For training the CASE method, we modify their code by applying the $\delta$ as their conditions and keep the same dimension of the latent $z$ as ours because there are only five skill categories in our setting. To train each evaluation discriminator $D^{i}_{test}$, we utilize the reference motions of the $i$-th skill as positive data and all the other reference motions as negative data. We use the same hyperparameters as those used in training the imitation policies; however, the discriminators for evaluation are independently trained for fair evaluation. During the skill evaluation, we collect 1 hour of ball tracking data of a match between two high-ranking players using the SPINSIGHT software\footnote {\url{https://spinsight.com/}} including the speed, including the position and speed of the ball when it contacts the paddle, touches the table, and passes the net. We use this ball tracking data to test the skill controller's ability with more challenging cases.

\section{Strategy evaluation}
For the opponent used in strategy evaluation, the \textit{random op} utilizes a strategy that uniformly samples the skill to use and the target land location of the ball. For the \textit{video op}, we collect 20 minutes of high-ranking player broadcast video and followed the annotation process of ~\cite{DBLP:journals/tog/ZhangYMGFPF23} to get the video expert demonstration $\{(o^\mathrm{video}_k, c^\mathrm{video}_k)\}^{K}_{k=1}$, by which the video strategy is trained.
% We utilize the same architecture as ours to train the video strategy for one iteration and fix it as the strategy for video op. 
% Using the demonstration, the video strategy is trained

We utilize reinforcement learning (RL) trained with PPO~\cite{DBLP:journals/corr/SchulmanWDRK17} as a baseline. We collect state action pairs $(s,\tilde{s},b,c)$ to update the policy. For the competition setting, the reward is defined as $r_g$, which is the task goal reward applied to the final step, $r_g = 10$ if the agent wins and $r_g = -10$ if it loses. Empirically, we find this sparse goal reward is actually better than the combination of $r_g$ and the continuous reward $r$ described in Section 4.2. For the cooperation setting, the reward is 1 for each time step and we set the episode length to $1000$. 

We also report the results of our strategy learning approach at each iteration. We report the winning rate for each iteration in Table~\ref{tab:wr_iter} and average rounds for each iteration in Table~\ref{tab:rally_iter}. We observe that the improvements in the winning rate at the first iteration are the most significant and then it converges gradually. For example, the winning rate increases by about 8\% in the first iteration and then only increases by about 10\% in the next four iterations. For the cooperation setting, the average rounds at the first iteration are close to that of the last iteration. We think the cooperation task is easier to learn compared to the competition match.

The video op incorporates real-world demonstrations, making them initially more challenging to defeat (48\% winning rate). However, with more iterations, the random op can sample a broader and more diverse set of strategy decisions. We can observe the final winning rates are similar. In addition, the more predictable nature of the video op makes cooperation easier. On the other hand, as an opponent, video op provides a relatively fixed challenge while random op provides more diverse and unpredictable challenges, which leads to our method being stronger when trained against a random op, as shown in Table 4 in the main paper.

% \jungdam{I don't understand what you want to say by using the two tables below? Aren't they just numbers RL vs. (random op, and video op)? Its a bit confusing for me because the experiment results appear abruptly. All the other paragraphs and sentences are about the details of experiments.}

\begin{table}[!h]
\caption{Winning rate for each iteration.}
\vspace{-0.1in}
\label{tab:wr_iter}
\begin{minipage}{\columnwidth}
\begin{center}
\begin{tabular}{lcccccc}
  \toprule
   Iteration & 0 & 1 & 2 & 3 & 4 & 5  \\ \midrule
Random op & 0.500 & 0.582 & 0.639 & 0.640 & 0.659 & 0.687  \\ 
Video op & 0.482 & 0.549 & 0.599 & 0.628 & 0.662 & 0.681   \\ 
  \bottomrule
\end{tabular}
\end{center}
\end{minipage}
\end{table}

\begin{table}[!h]
\caption{Average rounds for each iteration.}
\vspace{-0.1in}
\label{tab:rally_iter}
\begin{minipage}{\columnwidth}
\begin{center}
\begin{tabular}{lcccccc}
  \toprule
   Iteration & 0 & 1 & 2 & 3 & 4 & 5  \\ \midrule
Random op & 10.9 & 13.4 & 14.2 & 14.9 & 15.7 & 16.4  \\ 
Video op & 12.8 & 14.3 & 16.9 & 17.2 & 17.8 & 18.2   \\ 
  \bottomrule
\end{tabular}
\end{center}
\end{minipage}
\end{table}

\bibliographystyle{ACM-Reference-Format}
\bibliography{sample-bibliography}

%% file: Latex/1Introduction.tex
\section{Introduction}

The integration of deep learning into physics-based character animation has led to significant advancements in generating agile and natural motion, enhancing the lifelike quality of characters in complex environments. To increase the versatility of these characters, it is essential to ensure that their skills can be reused in environments or conditions that may not precisely match their training data. To achieve this goal, recent approaches have focused on learning reusable skill embeddings. These approaches are typically trained in two stages. Initially, characters learn various skill embeddings by imitating reference motions. Then, in the task training stage, they apply these skills to accomplish diverse tasks. These approaches have demonstrated remarkable success in generating natural motion in various environments.

However, when the differences between the skills are subtle, these approaches often suffer from mode collapse during the task training phase. Specifically, although agents (a.k.a. characters) can learn various skills during the imitation stage, they tend to use a limited set of skills for the downstream tasks, neglecting the diversity of their learned skills in the imitation stage. Thus, mode collapse restricts the agents' potential in scenarios that require a diverse set of skills. Mode collapse also restricts exploration during RL training, resulting in sub-optimal task performance. 

Another relatively unexplored topic relates to the decision strategy of agents, particularly their ability to dynamically formulate decision strategies that encompass skill selection and associated skill goals in response to task demands. Most previous studies either have not required a diverse skill set or have relied on a human user to manually determine skills for the agents. Agents have generally not been equipped with the capability to employ different strategies to adapt to complex and dynamic environments.

Our research introduces a learning approach to enhance both the skill and strategic decision-making capabilities of physically simulated agents. First, we develop a hierarchical skill controller that enables agents to utilize different table tennis skills and transition among them rapidly. This controller effectively addresses mode collapse during task training. Second, we develop a method for strategy learning, enabling agents to explicitly select and utilize specific skills for different types of interaction, whether competitive or cooperative. An overview of the results is in Figure~\ref{fig:teaser}. 

We demonstrate the effectiveness of our approach through two interaction environments: a table tennis match played between two simulated agents and a match between a human and a simulated agent in virtual reality (VR). In the agent-agent environment, the agents demonstrate improved skill diversity and decision strategy in simulated table tennis matches compared to results predicted by the previous techniques. In the human-agent interaction environment, we evaluate both cooperative and competitive scenarios in real-time interactions between humans and agents. These environments not only validate our approach but also provide platforms for future research into complex agent behaviors and human-agent dynamics. Code and data for this paper are at \url{https://jiashunwang.github.io/PhysicsPingPong/}.

We summarize the contributions of this paper as follows:
\begin{itemize}
\item  A hierarchical skill controller that empowers physically simulated agents to explicitly perform various skills, enabling rapid skill transitions. An interaction learning framework designed to create a decision strategy allows agents to continually learn and adapt, meeting the demands of competition or cooperation in dynamic environments with other agents and with humans.
\item Novel results demonstrating our learning framework's capacity to generate intelligent decisions and natural motions for table tennis in two scenarios: agent-agent interactions in a simulated environment and human-agent interactions in a VR environment. The agent-agent environment is a platform for developing and testing competitive and cooperative algorithms while the VR environment allows natural human-agent interactions. 
\end{itemize}

%% file: Latex/2RelatedWork.tex
\section{Related Work}
We review the closest related work in physics-based character animation with reusable skills and multi-character animation. We review studies on transitions among skills as we develop a method for skill selection and transition. We further discuss relevant research in human-agent interaction in VR. 

\subsection{Physics-based Character Animation}
Incorporating physical laws into character animation allows for the development of controllers that generate more realistic behaviors~\cite{DBLP:conf/siggraph/HodginsWBO95, DBLP:conf/siggraph/LaszloPF96}. Optimization techniques, such as trajectory optimization~\cite{DBLP:journals/tog/LasaMH10, DBLP:journals/tog/YinCBP08, DBLP:journals/tog/MordatchTP12} and sampling-based methods~\cite{DBLP:journals/tog/LiuYPSX10, DBLP:journals/tog/LiuPY16} have been widely explored. Recently, deep reinforcement learning (DRL) has been shown to substantially enhance control capabilities~\cite{DBLP:journals/tog/LiuH17, DBLP:journals/tog/PengBYP17}. Due to its flexibility and ease of use, DRL methods eliminate the need for designing complex objective functions while delivering outstanding results and have attracted significant research interest as a result. 

Data-driven methods have become prevalent in physics-based character animation studies since a DRL-based method was introduced by Peng et al.~\shortcite{DBLP:journals/tog/PengALP18}. The idea has been extended for handling larger datasets~\cite{DBLP:journals/tog/BergaminCHF19, DBLP:journals/tog/WonGH20} and for allowing recombination of existing state transitions~\cite{DBLP:journals/tog/PengMALK21}. Recently, much attention has been paid to reusable motor skills. The idea is to learn a latent space of reference motions and then to reuse the learnt space for downstream tasks. Various latent models have been studied such as encoder-decoders with autoregression~\cite{DBLP:conf/iclr/MerelHGAPWTH19, DBLP:journals/tog/WonGH21}, spherical embedding~\cite{DBLP:journals/tog/PengGHLF22, DBLP:conf/siggraph/TesslerKGMCP23, DBLP:conf/siggrapha/DouCFKW23}, conditional variational autoencoder (VAE)~\cite{DBLP:journals/tog/WonGH22, DBLP:journals/tog/YaoSCL22}, and vector-quantized VAE~\cite{DBLP:journals/tog/ZhuZLH23}. Some researchers have also proposed part-wise models to maximize reusability of reference motions~\cite{DBLP:conf/siggraph/BaeWLM023, DBLP:journals/tog/XuSZK23}.

Our system is designed for table tennis games, involving two players (i.e., agents). Two or more agents have been created primarily with kinematic approaches~\cite{DBLP:journals/tog/ShumKSY08, DBLP:journals/tvcg/ShumKY12, DBLP:conf/sca/LiuHP06, DBLP:journals/tvcg/KwonCPS08, DBLP:journals/tog/WamplerAHLP10}. There exist two recent approaches~\cite{DBLP:journals/tog/WonGH21, DBLP:journals/tog/ZhuZLH23} demonstrating examples of physically simulated boxing.
Zhang et al.~\shortcite{DBLP:journals/tog/ZhangYMGFPF23} build a system to learn tennis skills from broadcast videos and produce rallies with a mirrored opponent. In their approach, kinematics-based motion generation is utilized first, followed by physics-based tracking, relying on residual forces and extra arm control for successful strikes. Skill and target selection are not learned but rather performed manually or randomly to create a scene including two players.
In contrast, our method learns not only agile and precise motor control to strike the ball but also strategies to select skills and targets based on the movement of the opponent and the ball.

\subsection{Transition of skills}
Option-based methods~\cite{DBLP:journals/ai/SuttonPS99, DBLP:journals/ker/JainKP21, DBLP:conf/iclr/BagariaK20, DBLP:journals/corr/abs-1712-00004, DBLP:conf/nips/KonidarisB09} represent skills as \textit{options}, which are sequentially constructed, with each option's execution in the chain enabling the agent to execute the subsequent option. Lee et al.~\shortcite{DBLP:conf/iclr/LeeSSHL19} propose learning additional transition policies to connect primitive skills and introduce proximity predictors, which yield rewards based on proximity suitable for initial states for the next skill.  
One challenge of transitioning between different skills to chain long-horizon tasks is addressed by terminal state regularization~\cite{DBLP:conf/corl/LeeLAZ21}.
Behavior Trees are also a common method for planning the transition between different states~\cite{DBLP:conf/icra/MarzinottoCSO14, DBLP:conf/icra/FrenchWPZJ19, DBLP:conf/icra/ChengKP23}.
These methods achieve skill transitions by ensuring that the terminal state of the previous stage is close to the initial state of the next stage. While these methods work well for tasks that are not time-sensitive, table tennis, which involves high-speed movements and rapid responses, poses a challenge as players do not always hit the ball from a well-defined initial state.

\subsection{Human-agent interaction}

Research has focused on human sports training within VR~\cite{DBLP:conf/ismar/LiuWMK20, DBLP:journals/vr/PastelPCCSNSW23}. However, these studies often lack a physically simulated opponent. There are commercial games that allow people to interact with an agent in VR for sports activities, such as boxing, golf, and badminton. Eleven Table Tennis~\shortcite{Eleven-Table-Tennis} is a VR-based table tennis game similar to the one we have constructed, which enables a human to play with an agent. However, this agent is not simulated with full-body dynamics, rather it is simulated with only a floating head and a floating paddle. Advances in GPU-accelerated simulation and our control algorithm, enable us to create a physically-simulated agent with full-body dynamics that can play in real-time with humans. 
Another relevant area involves enhancing the agent's capabilities with \textit{human-in-the-loop} methodologies~\cite{DBLP:journals/rcim/LiZLPL22, DBLP:journals/corr/abs-2112-07774, DBLP:conf/humanoids/SeoHSBGSZ23, DBLP:conf/chi/WangBHSTG23} using extended reality. 
Our work differs from previous studies by bringing humans and agents into a unified environment allowing bidirectional physical interaction, where they can cooperate and compete. 

%% file: Latex/3Overview.tex
\begin{figure}[t]
    \centering
\includegraphics[width=1\columnwidth]{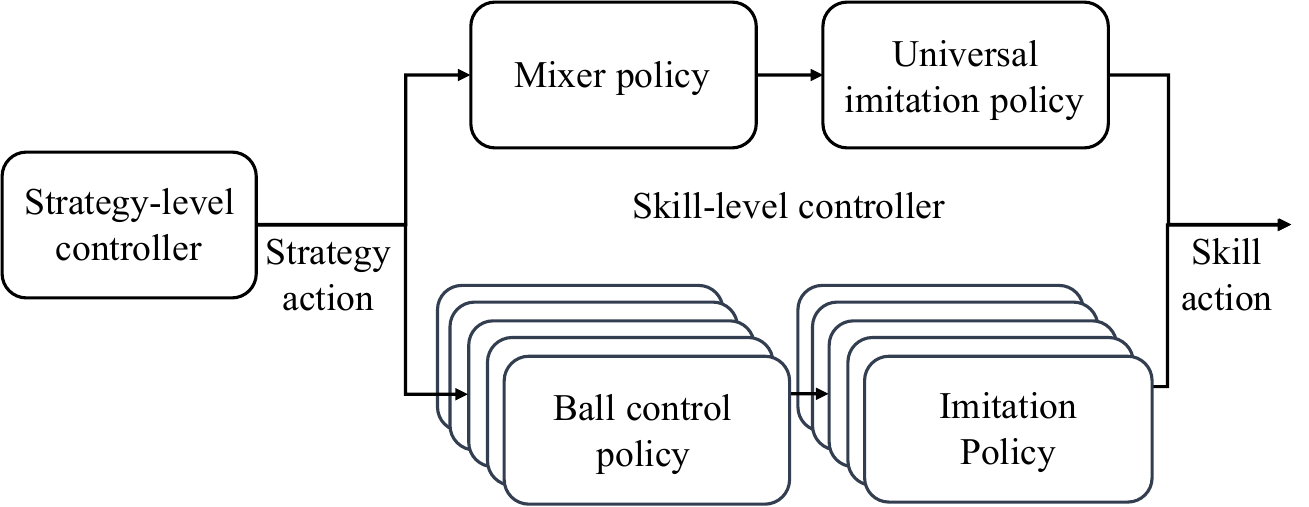}
\caption{An overview of our method. Strategy action includes the skill command and ball's target landing location. Skill action includes the target joint angles for PD controllers, blended from the outputs of imitation policies.}
    \label{fig:overview}
\end{figure}

\section{Method overview}

We propose a hierarchical approach that includes a strategy-level controller and a skill-level controller. The strategy-level controller takes the states of the agent, opponent, and ball as inputs, and outputs a strategy action, which includes the skill to use and the target landing location for the ball. Meanwhile, the skill-level controller takes the states of the agent and ball, along with the strategy action as inputs, and then generates a skill action, which includes the target joint angles for PD controllers. An overview of our method is in Figure~\ref{fig:overview} and Figure~\ref{fig:architecture} shows the architecture of our method.

%% file: Latex/4SkillLearning.tex
\begin{figure*}[t]
    \centering
\includegraphics[width=0.85\textwidth]{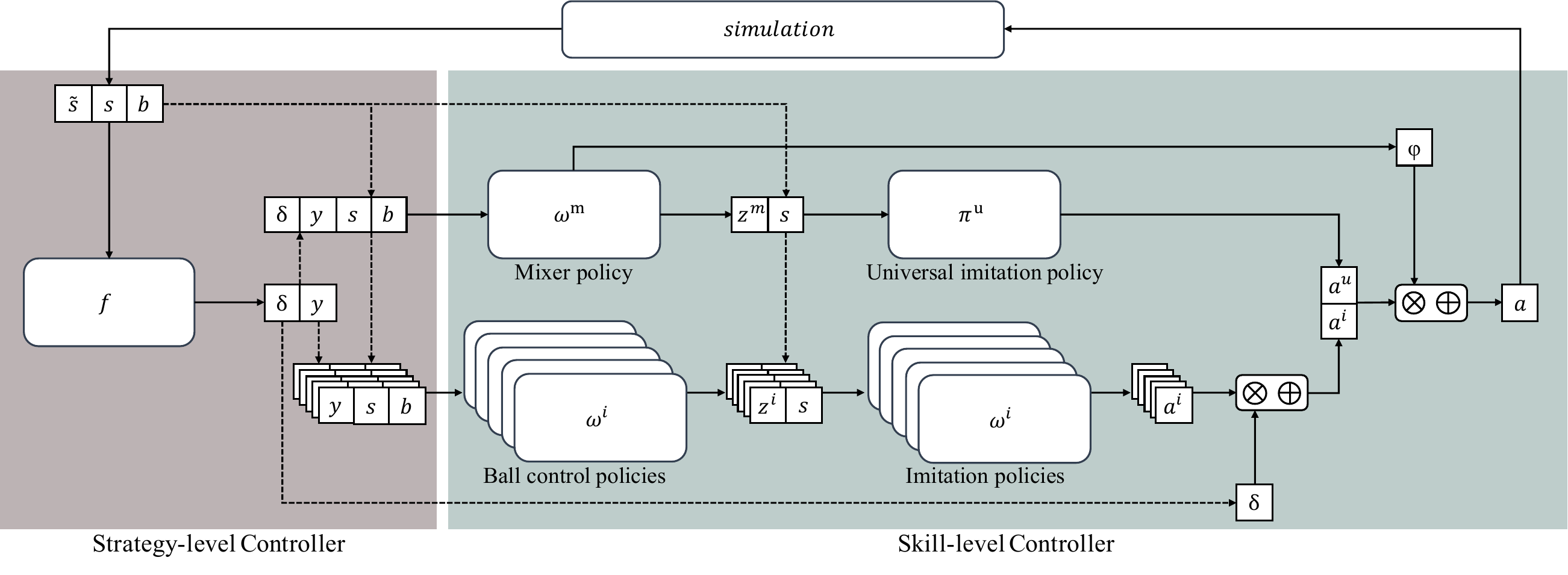}
\vspace{-0.1in}
    \caption{The architecture of our method. We train the skill-level controller through the stages of imitation policies, ball control policies, and finally, the mixer policy. We train the strategy-level controller after the skill-level controller is ready and its weight is frozen. $\otimes \oplus$ stands for the weighted sum in Equation~\ref{eq:mix}.}
    \label{fig:architecture}
\end{figure*}

\section{Skill-level Controller}
Three stages are required to train our skill-level controller. Initially, we train imitation policies using the motion capture data. Then the ball control policy for each skill is learned, which enables the agent to hit back balls using the corresponding imitation policy. Finally, we learn a policy that enables the agent to perform various skills sequentially while making plausible transitions among them. We call this policy the mixer policy. Once the skill-level controller is trained, the agent can proficiently and continuously execute various skills, sending balls to diverse target locations.

\subsection{Imitation Policy}
We first categorize the motion capture dataset into five subsets corresponding to each skill. This subdivision allows us to train the skill-specific imitation policies. We also utilize all the data to train a universal imitation policy. The imitation policy is represented as $\pi^{i}(a^i|s,z^i)$, where $i\in\{1, 2, 3, 4, 5, u\}$, $1\sim5$ are indices of different skills and $u$ is the index of the universal imitation policy. $z^i$ is a latent variable sampled from a hyper-sphere distribution, and $s$ is the agent's state. The goal of the imitation policy is to output an action $a^i$ that leads to simulated motions similar to the reference motions. Thus, each skill-specific imitation policy generates motions similar to its corresponding reference motion in each skill subset, while the universal imitation policy generates motions encompassing the entire motion capture dataset. When solving specific tasks in the later stage, using a single universal imitation policy trained with a variety of motions often leads to the mode collapse problem. The agent does not explore various available skills enough; instead, it repeats very limited skills, and the task performance remains sub-optimal. Our controller design is inspired by mixture-of-experts and mitigates this problem.
Each imitation policy $\pi^{i}(a^i|s,z^i)$ is built by the adversarial framework ASE~\cite{DBLP:journals/tog/PengGHLF22}, where the policy is updated so that it tricks a motion discriminator $D^{i}$. The transitions $d_{M^{i}}(s,s')$ existing in the motion capture dataset are used as positive samples while the transitions $d_{\pi^{i}}(s,s')$ generated from the policy $\pi^{i}$ are used as negative samples. The discriminator is trained by minimizing:
\begin{align}
    \mathop{\min}_{D^{i}}  & -\mathbb{E}_{d_{M^{i}}(s, s')} \log(D^{i}(s, s')) -\mathbb{E}_{d_{\pi^{i}}(s, s')} \log(1 - D^{i}(s, s')) \nonumber \\
    &+ \lambda_{gp}  \mathbb{E}_{d_{M^{i}}(s, s')} \left\|\nabla_{\phi}D^{i}(\phi)\big|_{\phi=(s,s')}\right\|^{2},
\end{align}
where the last term is a gradient penalty regularization with a constant factor $\lambda_{gp}$.
We train encoders $q^{i}$ to encourage correspondence between the transition $(s,s')$ and the latent variable $z^i$. The encoder is modeled as a von Mises-Fisher distribution and it is trained by maximizing its log-likelihood:
\begin{equation}
\begin{split}
    \mathop{\max}_{q^{i}}\mathbb{E}_{p(z^i)}\mathbb{E}_{d^{\pi^{i}}(s,s'|z^i)}[\log q^{i}(z^i|s,s')], \\
    q^{i}(z^i|s,s') = \frac{1}{Z}exp(\mu_{q^{i}}(s,s')^{T}z^i)        
\end{split}
\end{equation}
where $\mu_{q^{i}}(s,s')$ is the mean of the distribution, and Z is a normalization constant. Given a discriminator $D^{i}$, the reward to train $\pi^{i}$ is defined as:
\begin{equation}
    r_t = - \log  (1 - D^{i}(s_t,s_{t+1})) +\beta \log q^{i}(z^i_t|s_t,s_{t+1}).
\end{equation}
where $\beta$ is the relative weight. Additionally, a diversity term is included to encourage different latent variables to represent distinct motions. Bringing everything together, the objective of $\pi^{i}$ becomes:
\begin{align} 
\mathop{\max}_{\pi^{i}} \ & \mathbb{E}_{p(Z)} \mathbb{E}_{p(\tau|\pi^{i},Z)}[\sum_{t=0}^{T}\gamma^{t}(r_t)] \\
&- \lambda_{D^{i}}\mathbb{E}_{d^{\pi^{i}}(s)} \mathbb{E}_{z^i_1,z^i_2\sim p(z^i)}\left[\left(\frac{D_{KL}(\pi^{i}(\cdot|s,z^i_1),\pi^{i}(\cdot|s,z^i_2))}{0.5(1-z^i_1 z^i_2)}-1\right)^{2}\right], \nonumber 
\end{align}
where $D_{KL}(\cdot||\cdot)$ measures the KL-divergence between two distributions, $z^i_1$ and $z^i_2$ refer two different latent variables, $\gamma$ is the discount factor and $\lambda$ is a constant used to balance the weight.

\subsection{Ball Control Policy}
Once the agent can imitate each skill $i\in\{1, 2, 3, 4, 5\}$, we train ball control policies $\omega^{i}(z^i|s,b,y)$ to enable the agent to hit and move a ball launched from a random location to the desired location, where $s$ denotes the state of the agent, $b$ represents the state of the ball and $y$ is the target landing location for the ball.
The task reward $r$ is a composite of three terms: the paddle reward $r_p$, the ball reward $r_b$, and the style reward $r_{s}$,
\begin{equation}
r(t) = w_p r_p(t) + w_b r_b(t) + w_s r_s(t),
\end{equation}
where $w_p$, $w_b$, and $w_s$ are the relative weights. The paddle reward $r_p$ encourages the agent to position the paddle close to the ball. The reward is defined as:
\begin{equation}
r_p(t) = 
\begin{cases}
     &\exp(-4||x_p(t) - x_b(t)||^{2}), \text{ if } C_{bp}(t)=0, \\
     &0, \text{ otherwise},
\end{cases}
\end{equation}
where $x_p(t)$ and $x_b(t)$ represent the positions of paddle and ball, respectively, $C_{bp}(t)$ is a binary variable representing contact states. $C_{bp}(t) = 0$ means the ball has not contacted the paddle until time $t$, while $C_{bp}(t) = 1$ indicates the ball has contacted the paddle at time $t$ or contacted previously before time $t$. It will be reset to $0$ whenever the next ball is launched. 
% \jungdam{Add a sentence how this variable is reset.} 
The ball reward $r_b$ is given by:
\begin{equation}
r_b(t) = 
\begin{cases}
     &1 + \exp(-4||x_c(t) - x_t(t)||^{2}), \\
     &\qquad \text{ if } C_{bp}(t)=1 \text{ and } C_{bt}(t) = 0, \\
     &0, \text{ otherwise},
\end{cases}
\end{equation}
where $x_t(t)$ is the target landing location of the ball, $x_c(t)$ represents the anticipated landing location on the table, calculated using Newton's equation of motion for the point mass (i.e., quadratic trajectory), with its state corresponding to the current position and velocity of the ball. 
$C_{bt}(t)$ is a binary variable checking the contact history between the ball and the table, which is updated similarly to $C_{bp}(t)$.
The agent receives the maximum reward when it successfully hits the ball and the ball moves toward the target location.
We also apply the style reward, $r_{s} = -\log (1-D^{i}(s_t, s_{t+1}))$ in the task training similarly to ASE~\cite{DBLP:journals/tog/PengGHLF22}, where $D^{i}$ is the discriminator learned during the previous stage.

\subsection{Mixer Policy}
While our agent can play table tennis using the ball control policies with the corresponding imitation policies, its capability is limited to repeating a single skill. Simply transitioning from one controller to another during play often leads to failure due to a mismatch between the end state of one skill to the start state of the next.
To create plausible transitions among the different skills, we learn a mixer policy $\omega^{m}(z^m|s,b,\delta,y)$, which takes the agent state $s$, the ball state $b$, and the strategy action $(\delta, y)$ as input, where $\delta$ is a one-hot vector determining the skill to use and $y$ is the target ball landing location, then generates the latent variable for the universal imitation policy $\pi^{u}$ and a set of blending weights $\varphi$, mixing the skill actions in a joint-wise manner. 
In other words, $\varphi$ determines which policy the agent relies on among the transition and five different skills.
The target joint angles for PD controllers are computed as
\begin{equation}
\label{eq:mix}
\begin{split}
    a &= \varphi \odot \pi^{u}(\cdot|s, z^u) + (1-\varphi) \odot \sum_{i=1}^{5} \delta_i \pi^{i}(\cdot|s, z^i)
\end{split}
\end{equation}
where $\delta = (\delta_1, \delta_2, \delta_3, \delta_4, \delta_5)$ is a one-hot vector indicating the skill selected.
While training the mixer policy, the agent is asked to perform the ball control task with randomly launched balls, randomly selected skills, and random target locations. The same rewards used for learning ball control policies are employed, and the weights of all other policies remain frozen.

%% file: Latex/5StrategyLearning.tex
\section{Strategy-level Controller}
The strategy-level controller is developed by iterative behavior cloning inspired by~\cite{DBLP:conf/icml/OhGSL18}.
More specifically, we first collect interaction data by randomly sampling strategy actions during agent-agent play or human-agent interactions with VR. This data is then used to update the strategy-level controller, and we repeat this process by collecting new interaction data with the latest strategy-level controller. When collecting interaction data, there are two options: competition and cooperation. To train a competitive strategy, we choose data that results in victories, whereas in a cooperative strategy, we choose sequences where the opponent successfully catches the ball.

A strategy-level controller produces a skill index and a target landing location repeatedly so that they satisfy the requirements of different applications. More specifically, the strategy-level controller $f$ takes the strategy observation $o=(s, \tilde{s}, b)$ as input where $s$, $\tilde{s}$, and $b$ are the agent state, the opponent state, and the ball state, respectively, then outputs the strategy action $c=(\delta, y)$, where $\delta$ is a one-hot vector determining the skill to use, and $y$ is the target landing location of the ball. The strategy action is updated when the ball starts moving from the opponent to the agent.
\begin{algorithm}[t] 
\SetAlgoNoLine
\KwIn{Number of iterations $N$, interaction environment $Env$.}
\KwOut{Updated policy $f$.}
$f \leftarrow \mathrm{Random~initialization}$ \;
\For{$i \leftarrow 1$ \KwTo $N$}{
    $\{(o^\mathrm{expert}_k, c^\mathrm{expert}_k)\}^{K}_{k=1}\leftarrow$ Interact($Env$, $f$)\;
    Apply stochastic gradient descent to update $f$ using Equation~\ref{eq:bc}
}
\caption{Strategy learning}
\label{alg:bc}
\end{algorithm}
To effectively learn a strategy-level controller, we adopt a behavior cloning approach with iterative refinement, aiming to learn strategies from available expert demonstrations $\{(o^\mathrm{expert}_k, c^\mathrm{expert}_k)\}^{K}_{k=1}$ (see Algorithm~\ref{alg:bc}). As a structure of the controller, we utilize a Conditional Variational Autoencoder (CVAE) to model the stochastic nature inherent in sports gameplay. During training, the CVAE encoder takes $o$ and $c$ as inputs and generates the mean $\mu$ and variance $\sigma^{2}$ of the posterior Gaussian distribution $Q(u|\mu, \sigma^{2})$. We then sample a latent variable $u$ from this distribution and concatenate it with observation $o$ as input for the decoder, which reconstructs the action $c'$. The training loss is defined as:
\begin{equation}
\sum^{K}_{k=1} ||c^{\mathrm{expert}}_k-c'_k|| + \beta_{KL} D_{KL}(Q(u|\mu_k, \sigma^{2}_k)||\mathcal{N}(0,I)),
\label{eq:bc}
\end{equation}
where $D_{KL}(\cdot||\cdot)$ measures the KL divergence between the two distributions and $\beta_{KL}$ is the relative weight. During inference the decoder is utilized solely, it takes a randomly sampled latent variable $u$ and the observation $o$, and then generates the strategy action that guides the agent to perform a corresponding skill. If the opponent successfully returns the ball, this process repeats. 
We collect expert demonstrations from two different interaction environments ($Env$ in Algorithm~\ref{alg:bc}). The details of each environment will be explained in Section~\ref{sec:env}.

%% file: Latex/6InteractionEnv.tex
\section{Interaction Environment} \label{sec:env}
We introduce the agent-agent and human-agent interaction environments that we build to validate the strategy learning approach.

\textbf{The agent-agent interaction} environment is an environment where two virtual agents play table tennis with each other (Figure~\ref{fig:teaser} left column). We name one agent as \textit{our agent} and the other as \textit{the opponent}. In the process of learning a strategy-level controller for our agent, the opponent uses a fixed heuristic strategy-level controller while the controller for our agent is updated iteratively. More specifically, we let our agent the opponent play with each other using their own strategy-level controllers, collect those demonstrations, and then use them to update our agent's controller. If our goal is to learn a competitive strategy, that can beat the opponent, we selectively use demonstrations leading to wins. On the other hand, we use demonstrations where the opponent successfully returns the ball when aiming to learn a cooperative strategy. In our system, we utilize two types of heuristic strategy-level controllers: a random strategy and a video strategy. The random strategy selects skills and target landing locations randomly from a uniform distribution. The video strategy is constructed by using broadcast videos. We extract expert demonstrations $\{(o^\mathrm{video}_k, c^\mathrm{video}_k)\}^{K}_{k=1}$ from existing broadcast videos (20 minutes in total). Subsequently, we train a CVAE using the behavior cloning method.

\textbf{The human-agent interaction} environment allows a human user to play with a virtual agent. In our system, the user interacts with an agent by using a VR device, including a head-mount display and a hand controller ((Figure~\ref{fig:teaser} right column)). The VR interface operates through Unity while the physics-based simulation runs on Isaac Gym~\cite{DBLP:conf/nips/MakoviychukWGLS21}. 
To enable the simulated agent to interact with a human user, we physically simulate the user's paddle, with its position and orientation controlled via signals from the VR interface.
Specifically, for paddle control, we use the VR hand controller's Cartesian pose $q_\mathrm{user}$ and the simulated paddle pose $q_\mathrm{sim}$ to calculate the target velocity $\dot{q}_\mathrm{target} = (q_\mathrm{user} - q_\mathrm{sim}) / \Delta t$, where $\Delta t$ is the simulation step. 
We use this target velocity as an input to the velocity controller provided by the simulator.
For visualization, Unity takes the state of the simulated agent, user's paddle, and ball as inputs and renders them using visualization assets.
This implementation significantly reduces the amount of information exchange compared to a previous study that sent stereo images~\cite{DBLP:conf/humanoids/SeoHSBGSZ23}, enabling real-time interaction and gameplay. By considering a human user as the opponent, the strategy-level controller of the agent can be built through the same pipeline used for the agent-agent interaction environment.

%% file: Latex/7Experiments.tex
\begin{figure*}[t]
    \centering
\includegraphics[width=1.0\textwidth]{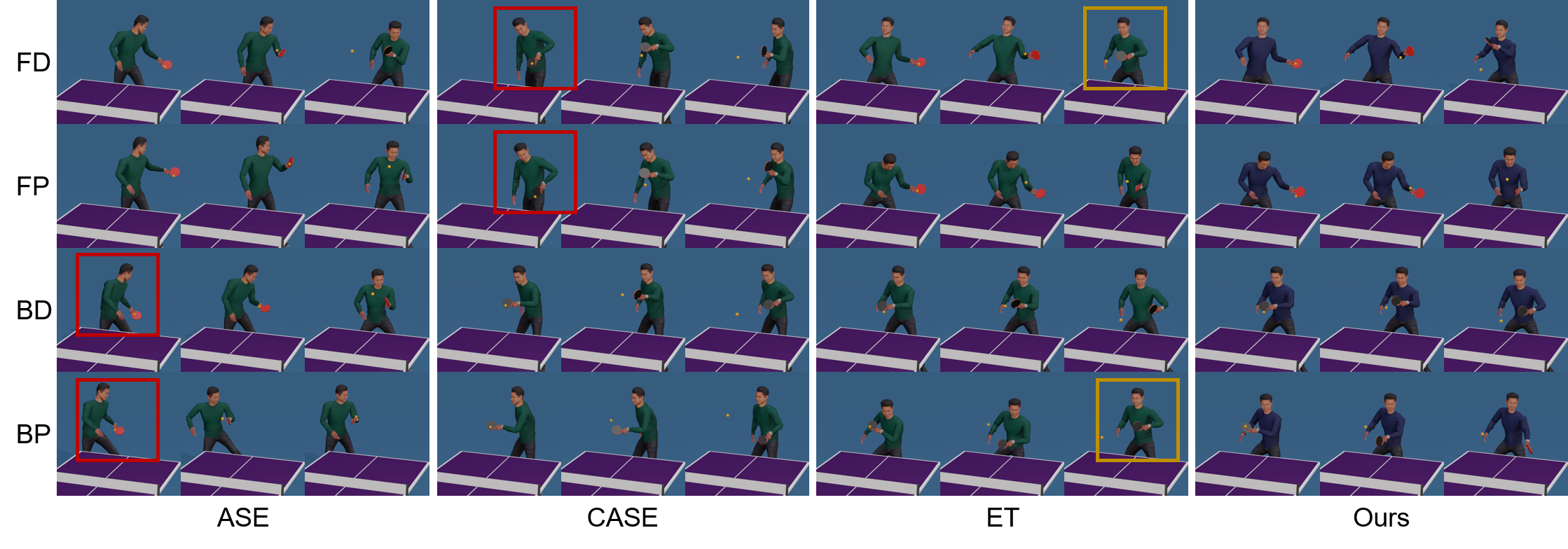}
\vspace{-0.3in}
    \caption{Comparison with other methods with four skill commands. ASE and CASE may use wrong skills as shown in the red box. ET may terminate earlier to return to a preparation pose, as shown in the yellow boxes.}
    \label{fig:compare}
\end{figure*}

\begin{figure*}[t]
    \centering
\includegraphics[width=1.0\textwidth]{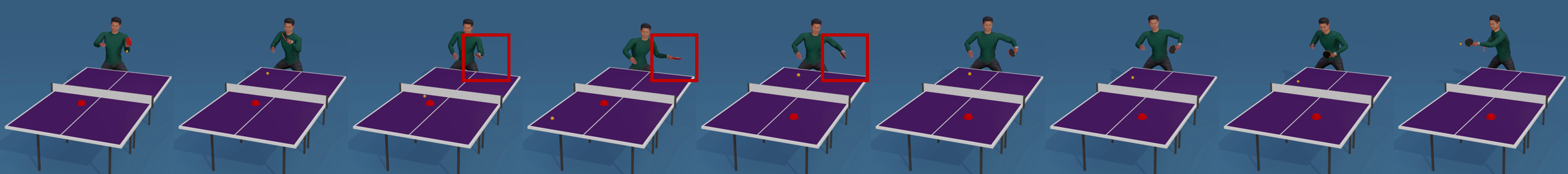}
    \caption{Transition results with only using forehand and backhand drive controllers. Both controllers are trained with random initialized configurations from the motion capture data. As shown in the red boxes, the agent attempts to use another forehand drive before the next ball is launched, which prevents it from switching back to a backhand drive in time.}
    \label{fig:random}
\end{figure*}

\section{Experiments}
We evaluate the skill-level controller based on motion quality and task performance. We assess the strategy-level controller by examining its effectiveness in agent-agent and human-agent interaction environments, with the demands of both competition and cooperation scenarios.

\subsection{Skill evaluation}
% The skill-level controller should be able to hit back balls launched randomly. 
We evaluate the skill performance from motion quality and task performance. 
The evaluation of motion quality measures the naturalness of generated motions when given the desired skill command and whether the agent performs the correct skill.
The evaluation of task performance measures the overall proficiency in playing table tennis. 
We compare our method with two state-of-the-art methods, ASE~\cite{DBLP:journals/tog/PengGHLF22} and CASE~\cite{DBLP:conf/siggrapha/DouCFKW23}, as well as an explicit transition model (ET) which is a variant of our method with the mixer policy $\omega^m$ removed from our model. We train an explicit controller to handle skill transitions by taking over the control when the ball passes the net until it is returned to the agent. The controller is also built using the ball control-imitation architecture. The key difference between our approach and ET is that ours provides continuous action blending with the selected skill's action at every time step, whereas ET does not.

\subsubsection{Motion quality} 
\label{sec:motion_quality}
We design three metrics to evaluate the motion quality, particularly to evaluate the naturalness and mode collapse. The first metric is \textit{Discriminator Score}, which measures how similar the current strike motion is to the reference motion of $i$-th target skill. Because we have five skills, we train a discriminator $D^i_{test}$ for each skill and utilize the following equation to calculate the score:
\begin{equation}
    \text{Discriminator Score i} = \frac{1}{T} \sum_{t=0}^{T-1}-\log (1-D^i_{test}(s_t, s_{t+1})),
\end{equation}
where $T$ is the length of a single strike motion. The details of training $D^i_{test}$ will be introduced in Appendix D. The second metric is \textit{Skill Accuracy}, to measure whether the agent performs the correct skill given the target skill command. Specifically, given a motion sequence, we first classify it by taking the index of the discriminator which provides the highest value.
Then, we compare it with the target skill command to calculate accuracy. The third metric, \textit{Diversity Score}, is designed to test whether motions for drive and push commands are distinctive enough. In table tennis, motions within each skill category (e.g., forehand drive vs. forehand push) might exhibit subtle differences, even though their roles in gameplay can be significantly distinct. \textit{Diversity Score} measures the capability of distinguishing motions that are visually similar. It is calculated by
\vspace{-0.05in}
\begin{equation}
\vspace{-0.05in}
    \text{Diversity Score} = \frac{1}{2N^{2}} \sum_{i \in \{1,3\}}  \sum_{m=1}^{N} \sum_{n=1}^{N} || s^{i}_{m} - s^{i+1}_{n}||,
\end{equation}
where $s^{i}$ is the state that the agent hits the ball under skill command $i$. Specifically $i \in \{1,3\}$ stands for forehand drive and backhand drive and $i+1 \in \{2,4\}$ stands for forehand push and backhand push respectively. 
$N$ is the total number of hits for each skill command. We only take into account the moment when the agent's paddle makes contact with the ball to calculate this score. 
The first two metrics evaluate a general skill mode collapse problem, for example, using forehand motions when being asked to use backhand. The third metric is specifically designed to measure if the agent has the ability to accurately perform drive and push skills.

The evaluation results are reported in Table~\ref{tab:motion quality}, where the values are computed with 10k balls randomly launched toward the agent equipped with the respective skill controller.
For the \textit{Discriminator Score}, our method significantly surpasses ASE and CASE, and achieves 15.6\% higher score than ET. These results prove our method generates motions that are the most similar to the reference target skill. 
As shown in the \textit{Skill Accuracy} results, our method uses the correct skills to hit the ball in most cases (0.76 in Table~\ref{tab:motion quality}). While ASE and CASE only use the correct skill with an accuracy of 0.38 and 0.47.  In \textit{Diversity Score}, our method achieves 30.7\%, 32.3\%, and 9.4\% higher scores than ASE, CASE, and ET respectively. We also show a qualitative comparison in Figure~\ref{fig:compare}. We find ASE and CASE often use forehand skills when asked to use backhand skills, or vice-versa, as shown in the red boxes in Figure~\ref{fig:compare}. And we can't observe any forehand smash skill. Even when the correct skills are used, the naturalness remains insufficient. ET often does not complete the skills; instead, the skills are terminated earlier to return to a preparation pose, as shown in the yellow box in Figure~\ref{fig:compare}. ASE and CASE often overlook skill commands, tending to use relatively fewer skills. This error occurs because, during the task training, these methods fall into mode collapse, making it challenging to effectively explore various skills. In contrast, our approach leverages an idea of the mixture-of-experts approach to avoid this problem. We further test the use of individual skill controllers without any design for transitions. Each skill controller is trained with randomly initialized configurations sampled from the motion capture data. As shown in Figure~\ref{fig:random}, after executing a forehand drive, the agent attempts another forehand drive before the next ball is launched—a typical behavior for single-skill controllers. This unnecessary movement prevents it from switching to a backhand drive in time, ultimately causing a missed shot.

\subsubsection{Task performance.} 
To assess the task performance of the skill controller, we evaluate two aspects: sustainability and accuracy. Sustainability is determined by the average number of successful continuous returns, while accuracy is measured by the average distance in meters between the target landing location and the actual contact location on the table. Besides testing on the training distribution, we collect some ball tracking data with faster ball trajectories from a match between high-ranking players and evaluate whether each method can perform well with the testset of the ball tracking data. We report the evaluation results in Table~\ref{tab:task performance}. The numbers in parentheses are the results of the fine-tuning experiments. Our method can achieve the largest number of average hits and the second-best accuracy. Although ET can achieve higher accuracy, it is not sustainable, especially for more challenging balls. It only achieves an average of 3.66 hits because it often lacks time to respond to the next ball due to the explicit transition design.

\begin{table}[t]
\caption{Comparisons on \textit{Discriminator Score}, \textit{Skill Accuracy}, and \textit{Diversity Score}. }
\label{tab:motion quality}
\begin{minipage}{\columnwidth}
\begin{center}
\vspace{-0.1in}
\begin{tabular}{lcccc}
  \toprule
   & ASE & CASE & ET & Ours \\ \midrule
  Discriminator Score & 1.62  & 2.28 & 4.95 & \textbf{5.72} \\
  Skill Accuracy & 0.38 & 0.47 & 0.69 & \textbf{0.76}\\
  % Diversity Score  & 8.48 & 8.42 & 8.32 & 8.69 \\
  Diversity Score  & 6.13 & 6.05 & 7.32 & \textbf{8.01} \\
  \bottomrule
\end{tabular}
\end{center}
\end{minipage}
\end{table}%

\begin{table}
\caption{Task performance evaluation. Our method can achieve the longest average hits and the second best accuracy.}
\label{tab:task performance}
\begin{minipage}{\columnwidth}
\begin{center}
\vspace{-0.1in}
\begin{tabular}{lcccc}
  \toprule
   & ASE & CASE & ET & Ours \\ \midrule
  Avg Hits   & 9.54 (5.94) & 8.79 (5.28 & 6.55 (3.66) &  \textbf{10.93 (6.28)} \\
  Avg error  & 0.28 (0.33) & 0.35 (0.39) & \textbf{0.25 (0.28)} & 0.26 (0.31) \\
  \bottomrule
\end{tabular}
\end{center}
\end{minipage}
\end{table}%

\subsubsection{Blending weights of the mixer policy.}
We test the agent with different skills to hit the ball and visualize the average blending weights $\varphi$ of the shoulder, elbow, and wrist joints in Figure~\ref{fig:weight}. We can observe that the weights of the mixer policy are usually lowest at the moment the paddle contacts the ball, and higher before and after transitions between different skills. It indicates a reliance on the pre-trained ball control policy during ball strikes, and on the mixer policy during transitions.

\begin{figure}[t]
\includegraphics[width=1.0\columnwidth]{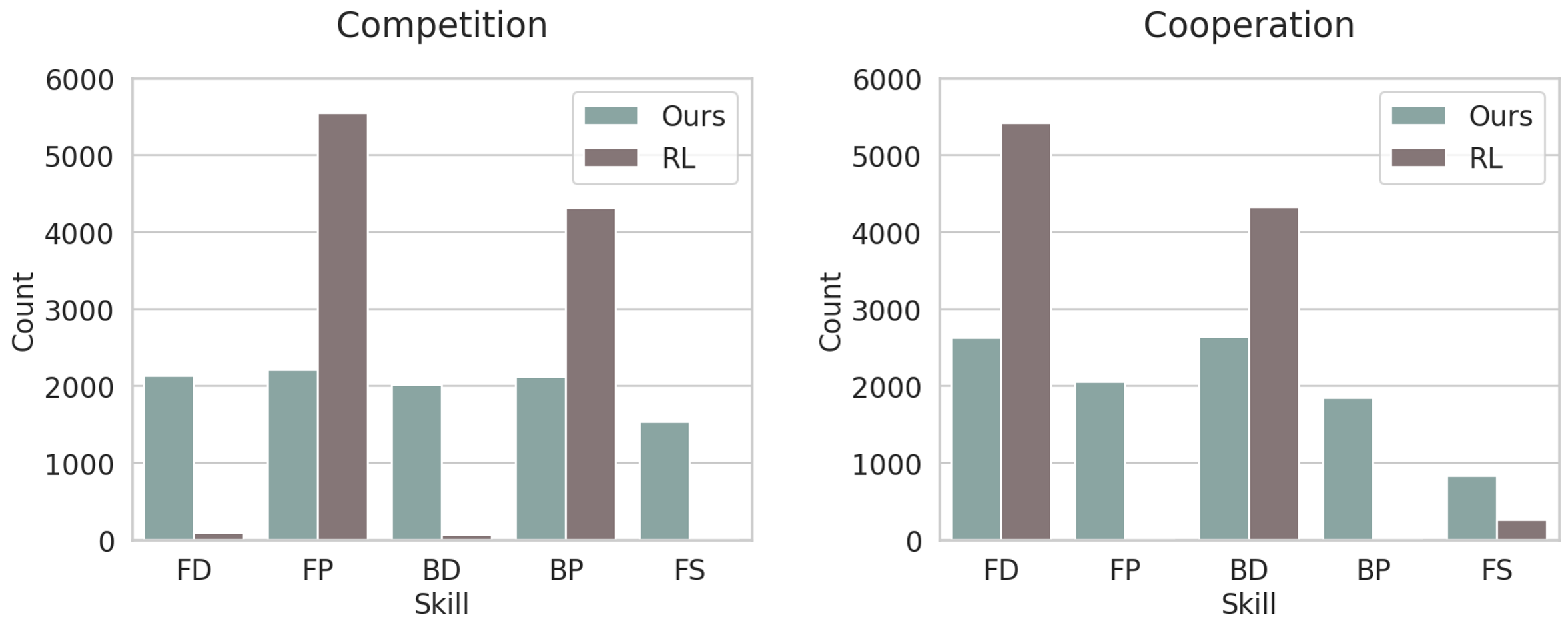}
\vspace{-0.28in}
  \caption{Skill command distribution of our method and RL.}
  \label{fig:his}
\end{figure}

\begin{figure}  \includegraphics[width=1.0\columnwidth]{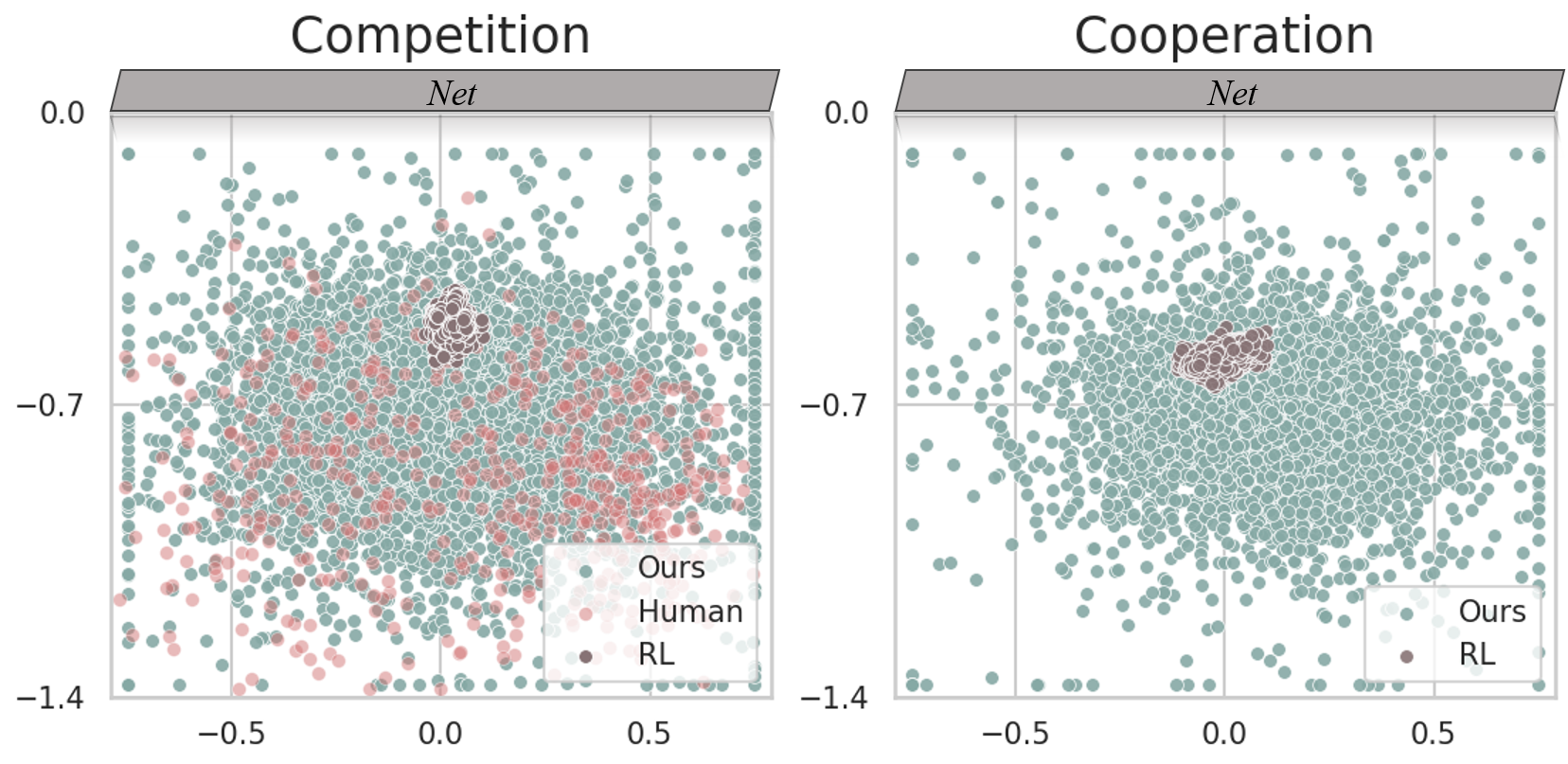}
  \vspace{-0.23in}
  \caption{Target landing locations of our method, RL and Human.}
  \label{fig:scatter}
\end{figure}

\subsection{Evaluation for agent-agent interaction}
We evaluate the performance of learned strategies in the agent-agent interaction environment under both competition and cooperation settings. The competition strategy aims to develop an agent that achieves a higher winning rate than the opponent. The cooperation strategy develops an agent that can play gently with the opponent to increase the length of rallies. As a baseline, we learn a strategy policy via reinforcement learning (RL). Please refer to Appendix E for details on training the RL baseline. 
Our method and the RL baseline are compared by having them play with two types of opponents: the random strategy and the video strategy opponent introduced in Section~\ref{sec:env}. Each evaluation is computed over 10k points.
Table~\ref{tab:strategy} shows the winning rate and the average rounds for the competition and cooperation settings. Our strategy learning algorithm can achieve higher winning rates for the competition setting and can maintain longer rallies for the cooperation setting for both opponents. 

In Figure~\ref{fig:his} and ~\ref{fig:scatter}, we visualize the histogram of skill commands from the strategy policies, and the target landing locations. In Figure~\ref{fig:scatter}, we also provide ball landing locations captured from real players during competitive matches. We observe that our method has a more similar distribution of landing locations to humans than RL. RL converges to less diverse skill commands and it only hits to a small region of the table. In contrast, our method utilizes various skills and target locations throughout the gameplay. We also include qualitative gameplay visualizations in Figure~\ref{fig:two_player}. We further let RL and our method compete with each other, and report the winning rates in Table~\ref{tab:rl_ours}. 
Each method has two strategy policies trained with two opponents, therefore, we have four matches in total. 
Because RL falls into a local minimum and overfits to a specific opponent, our method achieves a much higher winning rate.

\subsection{Evaluation of Human-agent interaction}
Before learning strategies for the human-agent interaction environment, we finetune the skill-level controller using the play data of a human user interacting with the agent equipped with the original skill-level controller. The finetuning is required because of the domain gap between what the simulated agent has experienced and the styles of a real human user in the VR environment. 
After finetuning the skill controller, strategies are learned by following similar procedures as the agent-agent interaction environment. 
For training a competition strategy, we use demonstrations that result in the agent winning points, thus presenting more challenging returns for the human opponent.
In contrast, for training a cooperative strategy, we use demonstrations where the human can maintain rallies, emphasizing easier ball returns for the human.
These demonstrations serve as expert demonstrations in Algorithm~\ref{alg:bc}.
% These collected sequences provide data to train the agent's strategy controller. 
We report the winning rate of the agent and the average hits between the user and the agent in Table~\ref{tab:human_agent}. When playing with the initial policies, the agent can achieve a winning rate of 64\%  and a rally with 4.04 hits on average. After two iterations of refinement of the competition strategy, the agent can achieve a winning rate of 78\% , and the average number of hits drops to 3.75. For the cooperative strategy, the winning rate drops to 58\%, and the user can achieve a rally with an average of 5.34 hits. These results demonstrate that our strategy learning algorithm is also effective for the human-agent interaction environment. We provide screenshots of real-time human-agent gameplay video in Figure~\ref{fig:vr}.

\begin{table}
\caption{Strategy evaluation. We report the winning rates for the competition setting and average rounds for the cooperation setting.}
\label{tab:strategy}
\begin{minipage}{\columnwidth}
\begin{center}
\vspace{-0.1in}
\begin{tabular}{lcccc}
  \toprule
  \multicolumn{1}{c}{} & \multicolumn{2}{c}{Competition}  & \multicolumn{2}{c}{Cooperation}   \\
  \cline{2-3} \cline{4-5}
   & RL & Ours  & RL & Ours\\ \midrule
  Random op & 0.641 & \textbf{0.687}   & 14.9 & \textbf{16.4}  \\
  Video op & 0.637 & \textbf{0.681}  & 15.6 &  \textbf{18.2} \\
  \bottomrule
\end{tabular}
\end{center}
\end{minipage}
\end{table}%

\begin{table}
\caption{Winning rates between our method and RL. The opponent in parentheses is the opponent during training of the strategy policy. }
\label{tab:rl_ours}
\begin{minipage}{\columnwidth}
\begin{center}
\vspace{-0.1in}
\begin{tabular}{lcc}
  \toprule
    & Ours (random op) & Ours (video op)  \\ \midrule
  RL (random op)   & 0.45 vs 0.55 & 0.47 vs 0.53 \\
  RL (video op) & 0.42 vs 0.58  &  0.42 vs 0.58 \\
  \bottomrule
\end{tabular}
\end{center}
\end{minipage}
\end{table}%

\begin{table}
\caption{Evaluation of human-agent interaction. }
\label{tab:human_agent}
\begin{minipage}{\columnwidth}
\begin{center}
\vspace{-0.1in}
\begin{tabular}{lccc}
  \toprule
   & Initial policies & Competition & Cooperation  \\ \midrule
  Winning rate   &  0.64
  & 0.78 & 0.58 \\
  Avg hits & 4.04 & 3.75  &  5.34 \\
  \bottomrule
\end{tabular}
\end{center}
\end{minipage}
\end{table}%

%% file: Latex/8Discussion.tex
\section{Discussion and Conclusion}

Although our method produces agents that play competitively and more naturally, it still has several limitations. First, although building individual policies for each skill and combining them via the mixer policy clearly improves the generated motion quality and task performance, our model would not scale well to a dataset including hundreds of different skills. Developing a hybrid model that combines our approach with a model learnable from unlabeled motions to achieve both high motion quality and scalability would be an interesting future research topic. 
Second, because our method is data-driven, the captured motion quality significantly affects the final motion quality. For example, the player tends to use large arm motions, and this motion style appears in our results as well. However, in matches, using less arm motion could be a way to conserve energy, and concealed movements can also confuse the opponent. 
Lastly, although we employ a rigid-body simulation for every component, including the ball, player, and table, where the ball can spin as well, air resistance is modeled using only damping based on velocity, rather than incorporating the Magnus effect, which bends the ball trajectory due to air pressure differences. This omission could impact the realism of our animations and the final strategies our system learned.

% \section{Conclusion}
In this paper, we introduce a learning approach for physics-based table tennis animation. We develop a hierarchical controller structure, which overcomes the mode collapse problem that appears frequently in reusable latent-based models. Our approach not only improves overall motion quality but also enables us to learn effective decision strategies for two types of environments: agent-agent and human-agent interactions. 

\begin{acks}
This work was partially completed during Jiashun Wang’s internship at The AI Institute. Jungdam Won was partially supported by Institute of Information \& communications Technology Planning \& Evaluation (IITP) grant funded by the Korea government(MSIT) [NO.2021-0-01343-004, Artificial Intelligence Graduate School Program (Seoul National University)] and ICT(Institute of Computer Technology) at Seoul National University. We would like to thank Murphy Wonsick for helping to build the VR system and Melanie Danver for rendering the results.
\end{acks}

%% file: Latex/FigureOnly.tex
% \begin{figure*}[t]
%     \centering
% \includegraphics[width=1.0\textwidth]{Image/compare.png}
%     \caption{Comparison with other methods with four skill commands. ASE and CASE may use wrong skills as shown in the red box. ET may terminate earlier to return to a preparation pose, as shown in the yellow boxes.}
%     \label{fig:compare}
% \end{figure*}

\begin{figure*}[t]
    \centering
\includegraphics[width=1.0\textwidth]{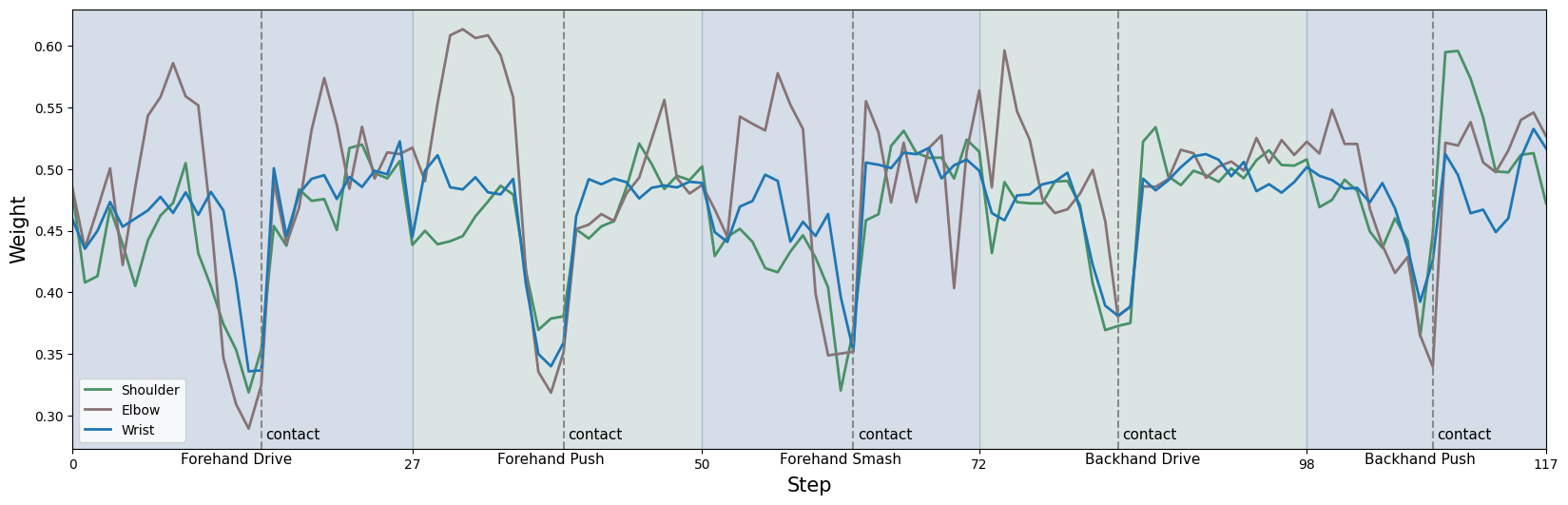}
\vspace{-0.3in}
    \caption{Visualization of the average blending weights $\varphi$ of the shoulder, elbow, and wrist joints. The weights of the mixer policy are usually lowest when the paddle contacts the ball, and higher before and after transitions between different skills.}
    \label{fig:weight}
\end{figure*}

\begin{figure*}[t]
    \centering
     % \vspace{0.in}
\includegraphics[width=1.0\textwidth]{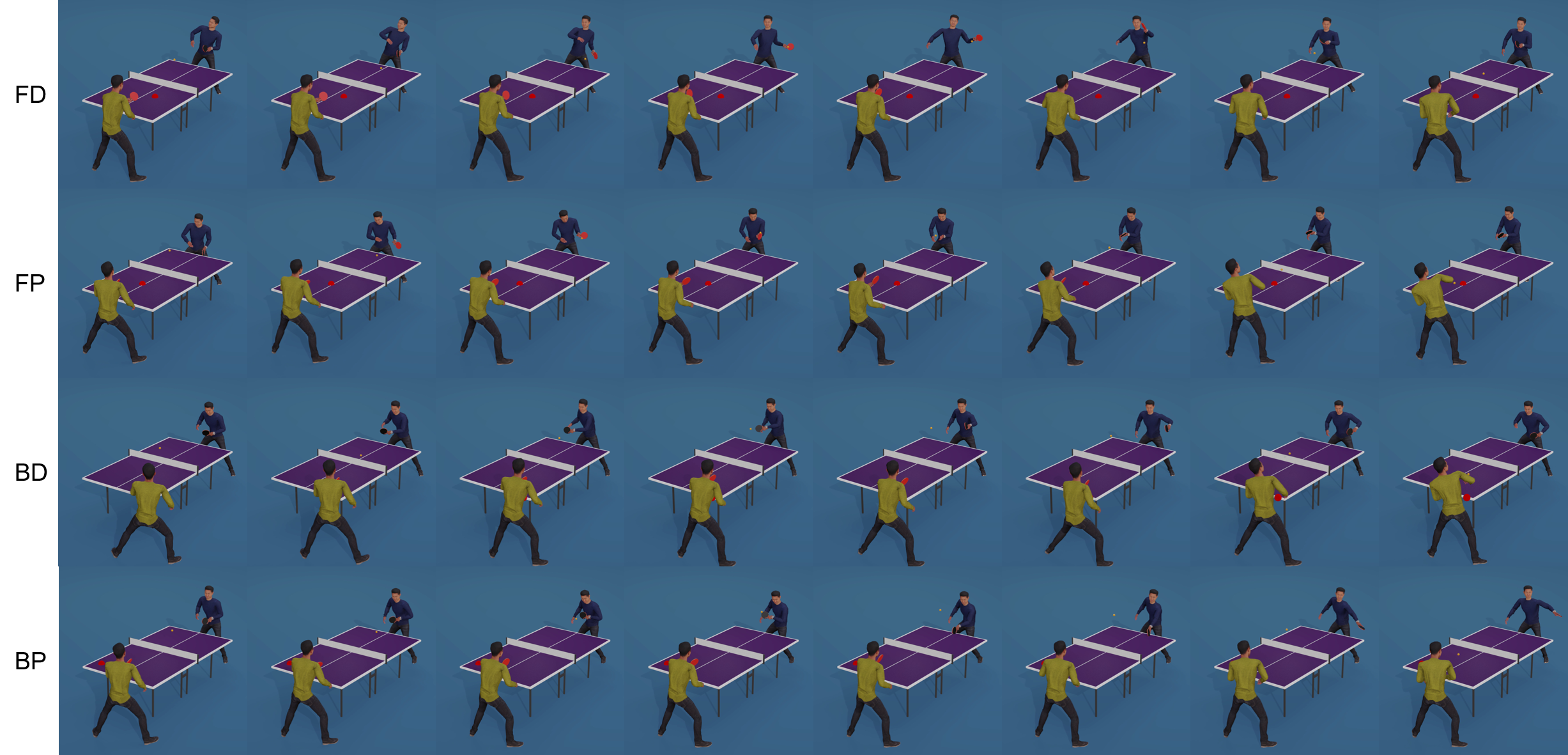}
% \vspace{-0.1in}
    \caption{Agent-agent gameplay. Blue agent is applying our strategy-level controller. The red dot is the target. We demonstrate four skills; the forehand smash is less obvious because the opponent does not deliver high and slow shots.}
    \label{fig:two_player}
\end{figure*}

\begin{figure*}[t]
    \centering
    \vspace{0.in}
\includegraphics[width=1.0\textwidth]{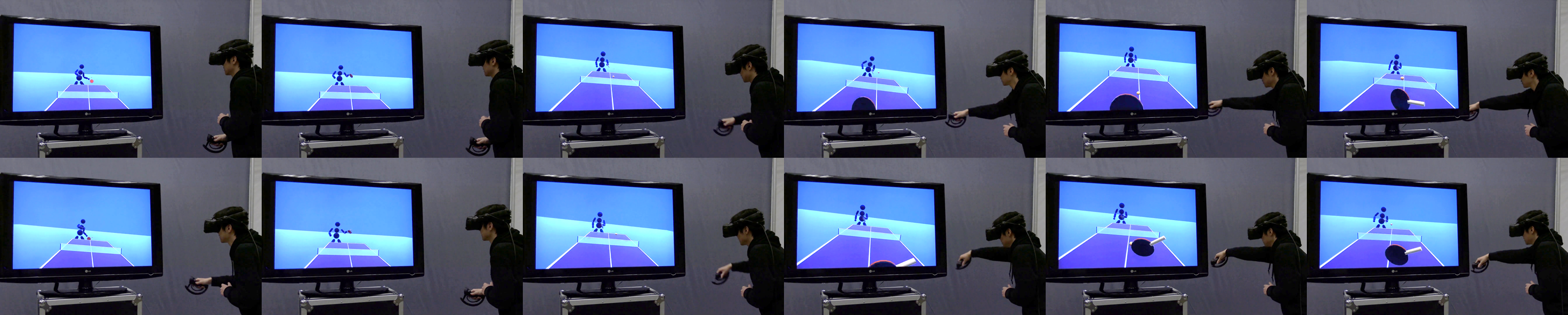}
% \vspace{-0.1in}
    \caption{Human-agent interaction screenshots. A human controls a simulated paddle and the agent is simulated and controlled by our method.}
    \label{fig:vr}
\end{figure*}